# Attached Decelerating Turbulent Boundary Layers over Riblets

Benjamin S. Savino[*1], Amirreza Rouhi[†2], and Wen Wu[‡1]
1. Department of Mechanical Engineering, University of Mississippi, University, MS, 38677, USA
2. Department of Engineering, Nottingham Trent University, Nottingham, NG1 4FQ, UK

**Turbulent boundary layers over riblets subjected to adverse pressure gradients (APGs) are investigated by direct numerical simulation. Multiple APG strengths and riblet sizes are examined, permitting evaluation of drag modification by riblets, and associated physical mechanisms, in various regimes established for zero-pressure-gradient (ZPG) riblet flows. The APG strengths are selected such that the flow remains attached. It is found that during APGs, riblets reduce drag beyond what has been achieved in ZPG flows. In extreme cases, an upstream force (i.e., negative drag) is attained. The significant drag reduction is found to be a product of Kelvin-Helmholtz roller vortices forming near the riblet crest, which are augmented in size, strength, and frequency during the APG. The preliminary results reported here indicate the need to modify existing metrics to predict drag reduction and the onset of KH rollers by riblets when the pressure gradient is non-negligible. Further analysis will be documented in the final paper.**

## I. Introduction

Riblets are one of the most extensively studied passive drag-reduction techniques, and have been shown to reduce friction drag by up to 10% in canonical turbulent flows (i.e., zero-pressure-gradient (ZPG) turbulent boundary layers (TBLs) and fully-developed turbulent channels) [1–15]. The fundamental shortcoming with the previous four decades of riblets research, however, is that emphasis has been placed on understanding their effect in equilibrium, ZPG flows. Meanwhile, in practice, they are employed to reduce drag in various aeronautical, maritime, and energy sectors. For example, Lufthansa and Aeroshark have begun applying riblet films to commercial airliners to investigate efficiency gains on an industrial scale [16].

Turbulent flows in such complex settings, however, rarely exhibit the well-behaved characteristics of ZPG TBLs, where riblets have typically been studied. A common non-equilibrium influence present in many practical applications is the pressure gradient (PG). PGs in external flows around complex geometries are caused by converging or diverging surfaces, a feature far more prevalent than flat plates in engineering contexts. For instance, flow over the suction side of an airfoil experiences a strong, spatially evolving adverse pressure gradient (APG), which significantly alters both the turbulence structure and the mean flow characteristics [17, 18]. Moreover, temporal variations in the driving force — such as those in pipe and closed-loop fluid systems, fluctuations in ambient or freestream velocity, or flow driven by reciprocating pumps — can also induce transient or sustained changes in the PG. It remains largely unknown how riblets interact with non-equilibrium, spatially-developing turbulence, and whether they maintain their drag-reducing properties when the flow is dominated by significant PGs. This hinders the effective use of riblets to reduce drag in realistic flow settings. Moreover, despite this uncertainty, reduced-order turbulence models used to represent the effects of riblets in engineering flows are often developed under the assumption that their behavior in zero-pressure-gradient (ZPG) flows carries over to spatially developing, PG-dominated flows [19, 20].

The limited understanding of riblet effectiveness in non-equilibrium flows motivates the present study. We focus on the APG in this paper. A few investigations have explored the influence of riblets in APG TBLs, both experimentally [21–23] and computationally [24, 25]. In general, these studies have shown that drag reduction tends to increase under APG conditions — relative to ZPG cases — with reductions of up to approximately 15%. However, the underlying physical mechanisms responsible for this enhanced drag reduction remain largely unexplored. Moreover, these studies focus on weak APGs, characterized by Clauser parameters ($\beta = (\delta^*/\tau_w)(dP/dx)$) on the order of $O(0.1–1)$, with little to no spatial variation. As a result, these weak APG TBLs differ little from canonical ZPG TBLs. The viscous-scaled peak-to-peak riblet spacing ($s^+ = s u_\tau/\nu$, where $u_\tau$ is the friction velocity and $\nu$ is the kinematic viscosity), a typical metric for gauging drag modulation in ZPG flows [2, 9], does not exhibit considerable variation. It always remains in the drag-reducing regime ($s^+ \in [0, 25]$). These features are not representative of many relevant flows, which as previously noted, experience significantly stronger APGs and exhibit considerable spatial variation. For example, on the suction side of an airfoil, $\beta$ can reach values on the order of $O(10^2)$ along the chord length even for an attached flow[26, 27].

---

[*]Ph.D. Candidate, AIAA Student Member
[†]Senior Lecturer
[‡]Assistant Professor. AIAA Member. Corresponding author. Email address: wu@olemiss.edu

The objective of this work is to investigate the influence of riblets in strong, spatially evolving APGs. The strong APG will result in significant spatial variation of $u_\tau$, which in turn varies the viscous-scaled riblet size across the domain. This, in conjunction with testing riblets of various physical sizes, allows the examination of riblet effectiveness in various regimes based on scaling laws developed in ZPG flows. These include: drag-reducing, drag-augmenting, and Kelvin-Helmholtz (KH)-roller-inducing (further discussion is provided in Sec. II(b)). This will allow us to: 1) characterize drag modification in these various regimes during flow deceleration; 2) determine if the quantitative metrics predicting drag decrease/increase from ZPG flows hold in APG flows; and 3) determine if the physical mechanisms responsible for drag modification in these regimes are the same in ZPGs and APGs. In the following abstract, we will first summarize the configuration and numerical methods in Section II, then discuss initial findings including boundary layer development, mean and instantaneous velocity fields, and preliminary turbulence statistics in Section III. The full paper will include a more detailed discussion regarding the mechanisms of the observations.

## II. Methodology

### A. Problem formulation

Turbulent boundary layers (TBLs) subjected to adverse pressure gradients (APGs) are simulated by direct numerical simulations (DNS). The APG is induced by imposing a decelerating freestream velocity profile along the top of the computational domain ($U_\infty(x)$, where $x$ is the streamwise direction). While some previous computational studies have applied a power-law distribution of $U_\infty(x)$ to achieve quasi-equilibrium decelerating boundary layers[28–30], the near-constant resultant Clauser parameter ($\beta$) is often small (typically $\leq O(1)$), indicating only a slight deviation from a conventional ZPG TBL and resulting in slow streamwise development. In the present study, a decreasing hyperbolic-tangent freestream velocity distribution is employed to induce a strong, spatially varying APG, thereby producing substantial variation in riblet spacing in wall units across the computational domain.

At the domain inlet, instantaneous velocity fields from an *a-priori* DNS of a ZPG TBL are applied. $\delta$ is the boundary layer thickness. The flow is allowed to develop under ZPG (*i.e.*, constant $U_\infty$) until it reaches $Re_\delta = 6800$. This streamwise location is hereafter referred to as the 'reference location', and all streamwise stations are defined with respect to it (*i.e.*, $Re_\delta = 6800$ at $x = 0$). Quantities at the reference location are indicated with the subscript '$o$'. For riblet cases, the riblets increase linearly in height over $1.5\delta_o$, reaching their maximum height at $x/\delta_o = -14$. This ensures negligible perturbation to the incoming smooth-wall TBL, and ensures the TBL is in an equilibrium state at the onset of the APG. The hyperbolic-tangent, decelerating $U_\infty(x)$ is applied downstream of the reference location. The corresponding wall-normal velocity is determined by mass conservation [31] and the spanwise velocity satisfies zero wall-normal gradient. The remaining boundary conditions are: no-slip on the bottom wall (riblet or smooth), periodic in the spanwise direction, and convective outflow.

Two levels of deceleration are examined, as shown in figure 1. The freestream velocity is decreased by 20% and 25%, respectively, over $30\delta_o$. The resulting Clauser parameter is shown in the middle panel of the figure. Here, the displacement thickness is calculated as:

$$\delta^*(x) = \int_0^{\delta(x)} \left(1 - \frac{U(x,y)}{U_e(x)}\right) dy, \tag{1}$$

where $\delta(x)$ is the local boundary layer thickness determined using the method described by Griffin *et al.* [32], $U_e(x)$ is the mean streamwise velocity at the boundary layer edge, and $y$ is the wall-normal direction. Our results show that when the bottom wall is smooth, the resulting peak values of $\beta$ are approximately 5 and 10, representing significantly stronger APGs than previously investigated for riblet-covered walls in Refs. [21–25]. The two sets of cases corresponding to these deceleration levels are denoted as $B5$ and $B10$, respectively. The friction velocity on the smooth wall (bottom panel in figure 1) decreases by ∼ 40% and ∼ 55% by the end of the domain in cases $B5$ and $B10$.

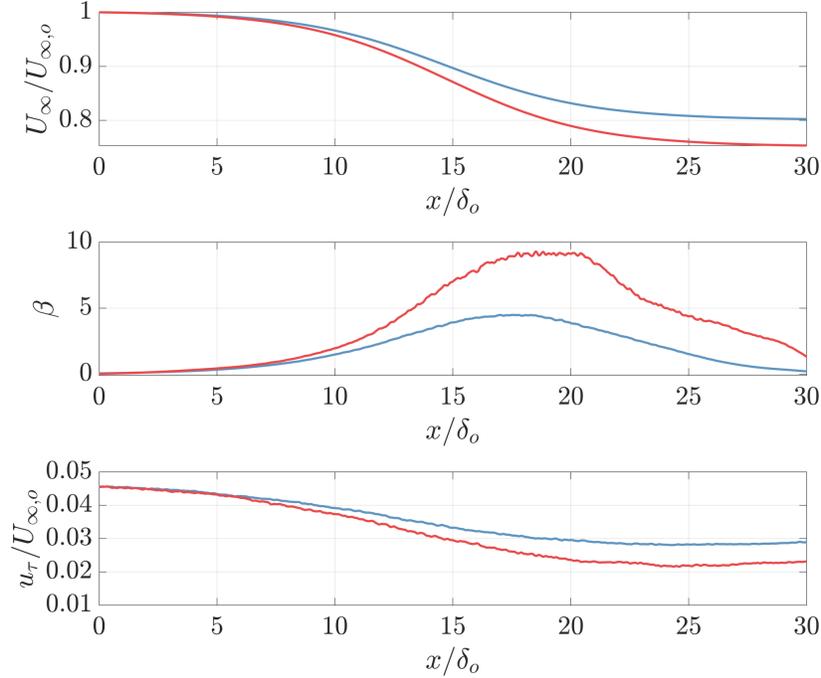

**Fig. 1 Top: freestream velocity profile applied at top boundary $U_\infty(x)$. Middle: Clauser parameter $\beta = (\delta^*/\tau_w)(dP/dx)$ over the smooth wall. Bottom: friction velocity over the smooth wall.** ——— B5; ——— B10.

### B. Riblet configuration

For each of the discussed APGs, streamwise-aligned riblets of various heights are applied on the bottom wall of the domain. The riblet shape in the spanwise direction ($z$) is defined by a sinusoidal function:

$$h_l(z) = 0.5h \left[ \sin\left(2\pi z/s\right) + 1.0 \right], \qquad (2)$$

where $h$ is the maximum riblet height measured from the lowest point in the valley and $s$ is the peak-to-peak spanwise spacing. In this study, the ratio $h/s$ is fixed to be $3/\pi$. A length scale is defined by the square root of the groove cross-section area (see figure 2), i.e., $l = \sqrt{\int h_l dz} = \sqrt{3/(2\pi)}s$. When normalized with the viscous length scale, $l^+ = lu_\tau/\nu$ has been shown to predict drag modulation by a variety of riblet shapes, exhibiting greater universality among geometries than $s^+$ [13–15].

Three riblet sizes are examined for each APG. They are $s_o^+ = 16$, 26, and 60 at the reference plane, corresponding to $s/\delta_o = 0.049$, 0.088, and 0.181. These yield $l_o^+ \approx 11$, 18, and 41. The smallest size corresponds to the optimal drag-reducing spacing widely reported in ZPG flows [2, 9, 13–15]. When $l^+$ approaches 20, drag penalty by riblets in ZPG flows begins, with the occurrence of KH roller vortices near the crest of the riblets (for certain riblet shapes, see figure 15(c) in the study by Endrikat *et al.* in Ref. [33], as well as Refs. [13, 14, 34]). For $l^+ > 40$, KH roller vortices are typically no longer observed in ZPG flows [33]. By selecting riblet sizes near these critical thresholds in the ZPG region preceding the APG, we aim to characterize potential changes in drag modulation, and the associated mechanisms, due to the deceleration. Specifically, since $u_\tau$ decreases under an APG, the $l_o^+ = 11$ case is expected to exhibit reduced drag reduction, with no KH rollers present throughout the domain. The $l_o^+ = 18$ case is anticipated to achieve greater drag reduction, with KH rollers showing near the reference plane and diminishing as $l^+$ decreases below the threshold of 20. For the $l_o^+ = 41$ case, a change from drag augmentation to drag reduction by the APG is expected, and although KH rollers are not anticipated at the reference location, they may develop downstream as $l^+$ further decreases under the influence of the APG.

Therefore, with the two examined levels of APG, a total of six riblet cases and two smooth-wall cases were performed. We name each case in the format of '$BXXlYY$', where $XX = 05$ corresponds to the APG with maximum $\beta \approx 5$, while $XX = 10$ corresponds to the APG with maximum $\beta \approx 10$. $YY = 00$, 10, 20, and 40 correspond to the smooth wall, and riblets with $l_o^+ \approx 11$, 18, and 41, respectively. Case parameters are given in Table 1.

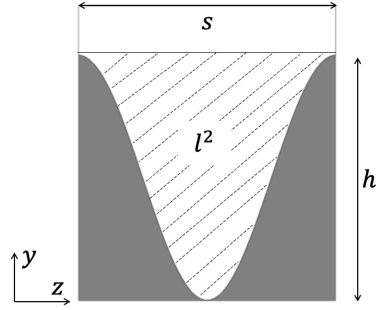

**Fig. 2   One period of the sinusoidal riblet profile with spanwise spacing ($s$), wall-normal height ($h$), and groove cross-sectional area ($l^2$) labeled.**

## C. Numerical methods

The incompressible Navier-Stokes equations

$$\nabla \cdot \mathbf{u} = 0; \quad \frac{\partial \mathbf{u}}{\partial t} + (\mathbf{u} \cdot \nabla)\mathbf{u} = -\nabla p + \nu \nabla^2 \mathbf{u} + \mathbf{f} \tag{3}$$

are solved by DNS. Here, $p$ is modified pressure, and $\mathbf{f}$ is the immersed boundary method (IBM) force used to enforce the no-slip boundary condition on the riblets. The IBM is based on the volume-of-fluid (VOF) approach [35, 36]. In the pre-processing stage, the fractional volume of each computational cell occupied by fluid is calculated. During the simulation, the predicted velocity at each time step is weighted by such a fraction and then corrected to be divergence-free via pressure projection, thereby enforcing the no-slip condition. The body force $\mathbf{f}$, which represents the force required to bring the fluid to rest within the solid region, accounts for all forces induced by the embedded structure. Because of the streamwise alignment of the riblets, only skin friction is present while form drag is zero. Wall-normal integration of $\mathbf{f}$ provides the total wall stress at each location. Readers are referred to Refs. [37, 38] for detailed proofs regarding the relationship between the IBM force and wall stress.

The streamwise ($L_x$), wall-normal ($L_y$), and spanwise ($L_z$) computational domain sizes for each case are provided in Table 1. $L_x$ is set to 1) allow the boundary layer to develop from the inlet Reynolds number to the target Reynolds number of $Re_{\delta,o} = 6800$; and 2) permit significant streamwise distance for the boundary layer to develop under the APG; 3) have the region of interest not impacted by the convective outflow. As mentioned above, the inflow is upstream of the reference plane at $x/\delta_o = -19.8$. $L_z$ is adjusted between cases to ensure exact periodicity of the riblet sinusoid. $L_y$ is set such that the freestream boundary on the top is far enough to allow the boundary layer to develop freely during the deceleration without significant blockage.

A Cartesian mesh is designed for each case to resolve the turbulent motions, including those on the sub-riblet scales. The $x$ grid is uniform throughout the majority of the domain, and linearly stretched near the outflow with a stretch ratio less than 3%. The $y$ grid is kept uniform below the riblet crest, with $n_h$ points resolving the riblet height. $n_h$ is varied between cases to ensure that $\Delta y^+_{(1)} \leq 0.6$ in all cases. Above the riblet crest, the $y$ grid is stretched using a hyperbolic-tangent profile with a stretch ratio less than 3%. The $z$ grid is uniform, with $n_s$ points resolving a riblet period. $n_s$ is adjusted between cases to ensure that the viscous scale of turbulence and the sub-riblet scale flow are resolved. Since $u_\tau$ decreases in the streamwise direction, the reference plane represents the location with the coarsest grid resolution in wall units. Nevertheless, the resolution achieved is comparable to, or better than, that of previous DNS studies on turbulent channel flows over riblets. The grid parameters are listed in Table 1.

The DNS is performed using a well-validated, in-house finite difference code [39–41], which solves the equations of motion on a staggered grid. Second-order-accurate central differences are used for all spatial derivatives. Time is advanced using a fractional step method [42]. Time is discretized semi-implicitly, with the wall-normal diffusion and the pressure gradient terms treated with the implicit Crank-Nicolson scheme, while all remaining terms are treated with an explicit Adams-Bashforth scheme. The Poisson equation is solved using a pseudo-spectral technique [43]. The code is parallelized using the message-passing interface (MPI) protocol.

Table 1   Simulation Parameters. See text for variable definitions.

| Case | $l_o^+$ | max ($\beta$) | $[L_x, L_y, L_z]/\delta_o$ | $[n_i, n_j, n_k]$ | Periods | $n_h$ | $n_s$ | $\Delta y_{1,o}^+$ | $\Delta x_o^+$ | $\Delta z_o^+$ |
|---|---|---|---|---|---|---|---|---|---|---|
| B05l00 | - | 5 | [55.3, 4.9, 4.9] | [1980, 270, 660] | - | - | - | 0.14 | 9.1 | 2.3 |
| B10l00 | - | 10 | [55.3, 4.9, 4.9] | [1980, 270, 660] | - | - | - | 0.14 | 9.1 | 2.3 |
| B05l10 | 11 | 5 | [57.7, 4.9, 4.8] | [2112, 296, 1584] | 99 | 54 | 16 | 0.13 | 8.8 | 0.9 |
| B10l10 | 11 | 10 | [57.7, 4.9, 4.8] | [2112, 296, 1584] | 99 | 54 | 16 | 0.13 | 8.8 | 0.9 |
| B05l20 | 18 | 5 | [56.0, 4.9, 4.8] | [2016, 270, 1512] | 54 | 54 | 28 | 0.47 | 8.9 | 0.95 |
| B10l20 | 18 | 10 | [56.0, 4.9, 4.8] | [2016, 270, 1512] | 54 | 54 | 28 | 0.47 | 8.9 | 0.95 |
| B05l40 | 41 | 5 | [58.3, 4.9, 4.9] | [2160, 309, 1080] | 27 | 100 | 40 | 0.58 | 9.9 | 1.5 |
| B10l40 | 41 | 10 | [58.3, 4.9, 4.9] | [2160, 309, 1080] | 27 | 100 | 40 | 0.58 | 9.9 | 1.5 |

### D. Data reduction

Statistics are collected after the flow in each case reaches a statistically steady state. For the smooth-wall cases, quantities are decomposed using the canonical Reynolds decomposition: $\phi(x, y, z, t) = \langle \overline{\phi} \rangle (x, y) + \phi'(x, y, z, t)$, where the operators $\overline{(.)}$ and $\langle . \rangle$ denote the temporal and spanwise average, respectively. $(.)'$ denotes the stochastic turbulent fluctuations. For the riblet cases, instantaneous quantities are first decomposed with a dispersive field, obtained by averaging in time and ensemble averaging over repeated riblet periods. Then, together with the local stochastic turbulent fluctuation as the residual, the decomposition reads: $\phi(x, y, z, t) = \overline{\phi}(x, y, z_r) + \phi'(x, y, z, t)$. Here, $z_r = \mathrm{mod}(z, s) \in [0, s)$ is the spanwise location within a riblet period corresponding to the global $z$ location. The dispersive field is then decomposed into a spanwise average over the riblet and the spatial variation about this average: $\overline{\phi}(x, y, z_r) = \langle \overline{\phi} \rangle (x, y) + \tilde{\phi}(x, y, z_r)$. This yields the 'double average' widely employed in rough- and riblet-wall turbulence literature [15, 33, 44–46]:

$$\phi(x, y, z, t) = \langle \overline{\phi} \rangle (x, y) + \tilde{\phi}(x, y, z_r) + \phi'(x, y, z, t). \tag{4}$$

For the operators introduced by this decomposition, those without a subscript denotes the intrinsic spanwise average over the fluid domain only, whereas $\langle . \rangle_s$ indicates the superficial spanwise average over the fluid and solid domains. Throughout this paper, temporally and spanwise ensemble-averaged primary variables are represented by capital letters as well.

## III. Results

### A. Boundary layer development

The mean velocity is compared in figure 3. The boundary layer thickness is indicated by the white dashed line, and the riblet crest is represented by the black horizontal line. As previously mentioned, the intrinsic spanwise average is taken in the spanwise direction, considering only the contribution from the fluid domain. It can be seen that the APG decelerates and thickens the boundary layer in all cases, though the extent of thickening varies between them. This is quantified by $\beta$ and $Re_\delta = \delta U_e / \nu$ in the top two rows of figure 4. The maximum value of $\beta$ increases dramatically with riblet size. For the largest riblet under the moderate APG (case B05l40), and the two largest riblets under the strong APG (B10l20, B10l40), $\beta$ approaches infinity shortly after the APG takes into action. This trend reflects a rapid deceleration near the wall (i.e., within the riblet region). However, the deceleration further away from the wall remains largely unaffected, resulting in negligible increases in $Re_\delta$ for all moderate-APG cases and the high-APG cases with small or medium riblets.

For two-dimensional separating flow over smooth walls, infinite $\beta$, resulting from wall shear stress approaching zero, serves as a necessary and sufficient indicator of flow separation. However, this criterion does not apply to surfaces with microstructures. Our previous study on separating flows over sandgrain roughness [47] demonstrated that reverse flow can occur within the roughness layer, producing zero or even negative wall shear stress well before separation occurs in the outer flow. A similar phenomenon is observed in the present study. The mean velocity contours clearly show the formation of a mean reverse flow within the riblet grooves, most prominently in cases B05l20, B05l40, B10l20, and B10l40. The wall shear stress and mean velocity at the riblet crest are shown in the bottom two rows of figure 4.

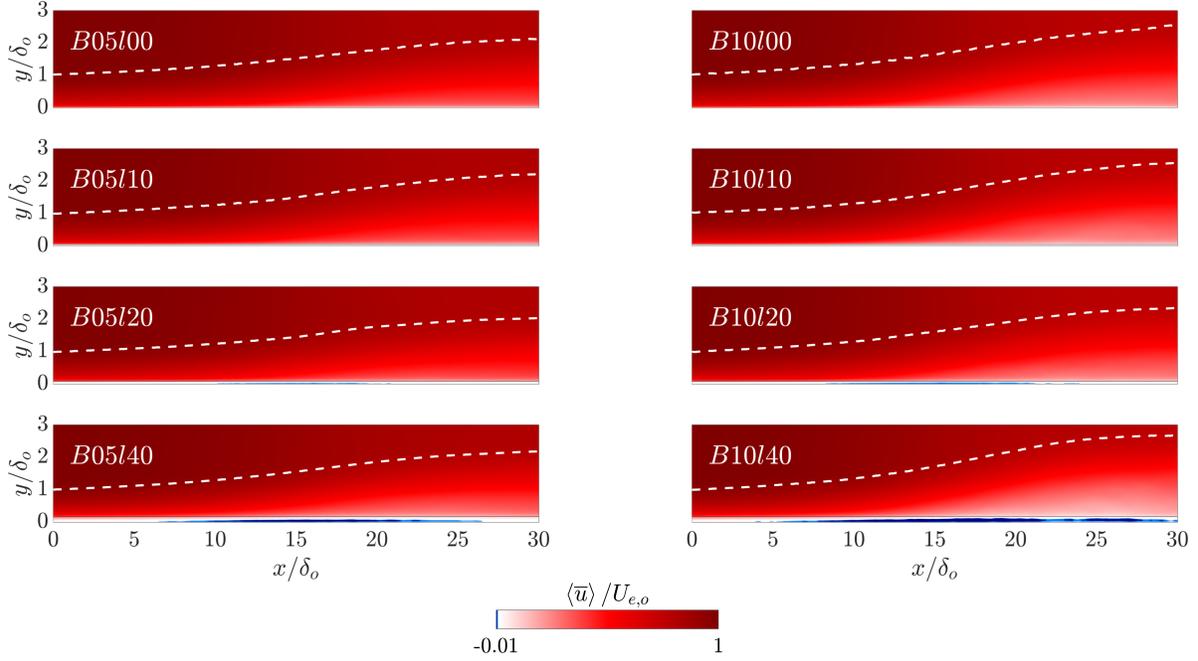

**Fig. 3** Contours of the time- and spanwise-averaged streamwise velocity. The superimposed white dashed lines denote $\delta$, while the thin horizontal black line denotes the riblet crest. The spanwise average is intrinsic (*i.e.*, over the fluid domain only) below the riblet crest. Note that the blue (reverse flow) and red (forward flow) regions are saturated at different magnitudes, with white representing zero velocity as the pivot.

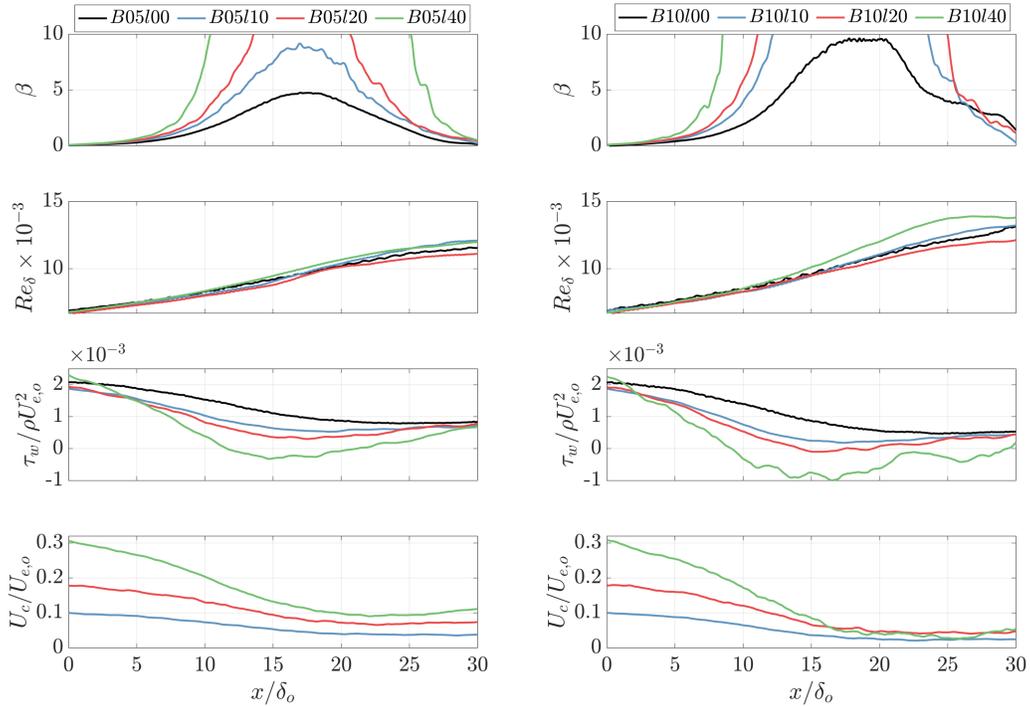

**Fig. 4** Streamwise development of various boundary layer parameters. From top to bottom: Clauser parameter ($\beta = (\delta^*/\tau_w)(d\langle \overline{p} \rangle/dx)$); $Re_\delta = \delta U_e/\nu$; wall shear stress ($\tau_w = \nu(\partial \langle \overline{u} \rangle/\partial y)$ in the smooth wall cases, see text for method of calculation in the riblet cases); mean streamwise velocity at the riblet crest ($U_c/U_{e,o}$). The left column compares the cases at the lower APG and the right column at the higher APG. In all plots, ( ——— ) $l00$; ( ——— ) $l10$; ( ——— ) $l20$; ( ——— ) $l40$.

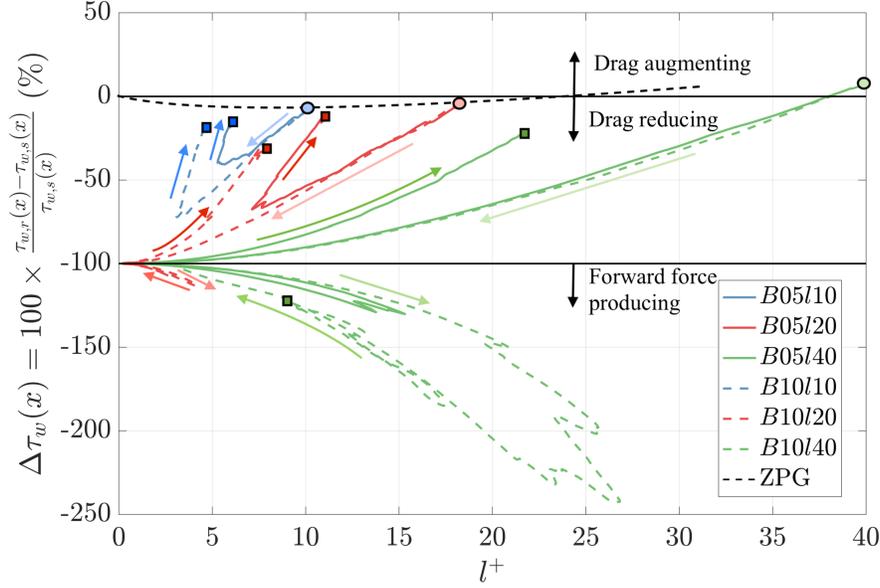

**Fig. 5   Drag curve for APG cases compared with the ZPG one. The arrows along the APG drag curves indicate the downstream direction, while the darkening shades of color represent locations further downstream. The beginning and end of the region of interest ($x = 0$ and $x/\delta_o = 30$) are indicated by circular and square markers, respectively. Drag augmenting, drag reducing, and forward force producing regimes are indicated.**

It can be seen that three cases - $B05l40$, $B10l20$, and $B10l40$ - exhibit negative wall shear stress under the imposed deceleration. However, none of the cases display a negative mean velocity at the riblet crest. Therefore, in all cases, the flow continues downstream along the riblet crests (i.e., no flow separation), while reverse flow develops within the riblet grooves. As we will explain momentarily, this mean reverse flow is due to the passage of consecutive vortices.

### B. Drag reduction quantification

As our primary focus, we quantify the drag change by comparing the smooth-wall and riblet cases. Note that drag reduction should be evaluated by comparing two flows that either have the same mass flow rate but differ in wall shear stress, or share the same wall shear stress but differ in flow rate. In the current study, $Re_\delta$ remains nearly the same between the smooth and riblet cases at each streamwise location in most cases. Therefore, we compare wall shear stress values at the same $x$ position to evaluate drag differences.

In ZPG flows, the drag modulation by riblets is predicted by the 'drag curve', which plots $\Delta \tau_w$ versus viscous-scaled riblet size, typically $l^+$ (e.g., figure 5 in Ref. [14]). $\Delta \tau_w$ is defined as:

$$\Delta \tau_w(x) = 100 \times \frac{\tau_{w,r}(x) - \tau_{w,s}(x)}{\tau_{w,s}(x)} \%, \qquad (5)$$

where subscripts 'r' and 's' denote the riblet and smooth wall, respectively. Figure 5 presents our results alongside those from ZPG flows. Striking differences, highlighting the non-equilibrium effects, are clearly shown. The circular marker at the beginning of each of the plot is at the reference plane, while the square indicates $x/\delta_o = 30$. The $B05$ cases are shown with the solid lines, while the $B10$ cases are shown with dashed lines. The colors representing the riblet size at $x = 0$ are the same as figure 4. The drag curve from ZPG flows is given by the dashed black line.

Based on the drag curve of the ZPG fows, we are expecting to see the following changes: the $l10$ cases are designed to be near the optimum of the drag curve, and thus should transit leftward into the viscous, linear regime as $u_\tau$ (and thus $l^+$) decreases by action of the APG. While they should still be drag reducing, drag will be reduced by less than at the inflow. Similarly, the $l20$ cases should transit from the rightmost end of the drag-reducing regime toward the optimum (i.e., drag will be reduced by a greater percentage downstream of the inlet), while the $l40$ cases should transit from the drag-increasing regime toward the drag-reducing regime. These trends are expected to be enhanced at the higher APG.

At the reference plane in all cases, where the flow is under ZPG conditions, the drag modulation is as expected: the $l10$ and $l20$ cases decrease $\tau_w$ by ∼ 6%, while the $l40$ cases display a drag increase. However, none of the anticipated changes associated with the APG-modulated $l^+$ occur. As the APG becomes appreciable, $l^+$ decreases considerably. In all cases, this comes with a multifold augmented drag reduction compared to the ZPG region, and there is significant deviation from the ZPG drag curve (shown by the faint arrows in the figure). The prediction for the $l10$ cases that $\Delta\tau_w$ would increase does not hold. Furthermore, while greater drag reduction was expected for the $l20$ and $l40$ cases due to the faster reduction in $l^+$, $\Delta\tau_w$ appears to decrease with $l^+$ significantly faster than predicted by the ZPG drag curve. Note that augmented drag reduction by riblets in very weak APGs relative to ZPG has been reported [22]–[24], however $\Delta\tau_w$ was never reported to exceed $-15\%$. The observed deviation from the ZPG drag curve, and range of $45-250\%$ drag decrease, indicates that the current mechanisms responsible for drag modulation differ from those in ZPG and near-equilibrium APG TBLs. Therefore, corrections to the present ZPG metrics are needed to predict the riblet influence on skin friction in the present spatially developing, non-equilibrium APG TBL.

For the moderate APG ($B05$), the minimum realized $\Delta\tau_w$ decreases as riblet size increases, implying that drag reduction increases with riblet size. For cases $B05l10$ and $B05l20$, where $\tau_w$ remains positive, $\Delta\tau_w$ decreases with $l^+$ to a local minimum, after which it begins to increase toward the ZPG drag curve. It was found that the local minimum on the drag curve for these cases corresponds to the streamwise location where the APG begins to diminish, and the flow begins recovery toward an equilibrium ZPG TBL ($x/\delta_o \approx 25$). For the largest riblet size (case $B05l40$), where mean $\tau_w$ exhibits negative values, $\Delta\tau_w \to -100\%$ where $\tau_{w,r} \to 0$. A region of forward force, or thrust, is indicated by $\Delta\tau_w < -100\%$ where $\tau_{w,r} < 0$. The peak forward force was found to occur at the same streamwise location as maximum $\beta$: $x/\delta_o \approx 18$. $l^+$ resulting from the reverse flow in such region is about 15. Such a regime of the drag curve has not been reported, to the best of our knowledge. As the APG diminishes, $\Delta\tau_w$ returns to $-100\%$ and begins recovery toward the ZPG drag curve. The trend of the plots towards the end of the computational domain indicates that if given sufficient streamwise recovery distance under new ZPG conditions after deceleration, the wall stress may recover to the ZPG drag curve.

The qualitative behaviors are similar for the strong APG ($B10$, dashed lines) as the moderate APG ($B05$, solid lines). Interestingly, the decrease of $\Delta\tau_w$ with respect to $l^+$ during the APG onset occurs at the same rate (i.e., $d\Delta\tau_w/dl^+ \approx$ constant, and the solid and dashed lines collapse before they reach their individual minimum) for a given riblet size. As the APG increases, the curve approaches the point of total drag elimination more closely. This trend is not observed for the largest riblet tested, as both APG levels produce a net forward force. The key difference between the two APG levels in these cases is that the higher APG induces an increased forward force, resulting from a stronger mean reverse flow.

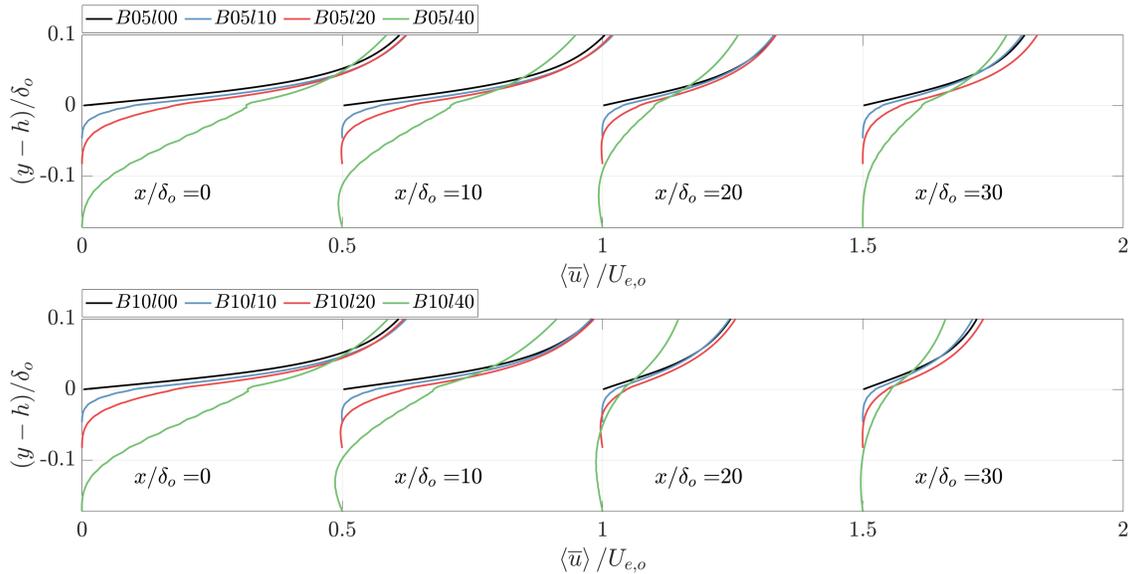

**Fig. 6** Mean streamwise velocity profiles extracted at $x/\delta_o = 0$, $10$, $20$, and $30$. The top plot displayed the $B5$ cases, and the bottom displayed the $B10$ cases. Color indicates riblet size, as in figure 4. The vertical coordinate is shifted by $1.0h$ for the riblet case such that the riblet crest corresponds to $(y - h) = 0$. Profiles are shifted right by $0.5$ units at each $x$ location for clarity.

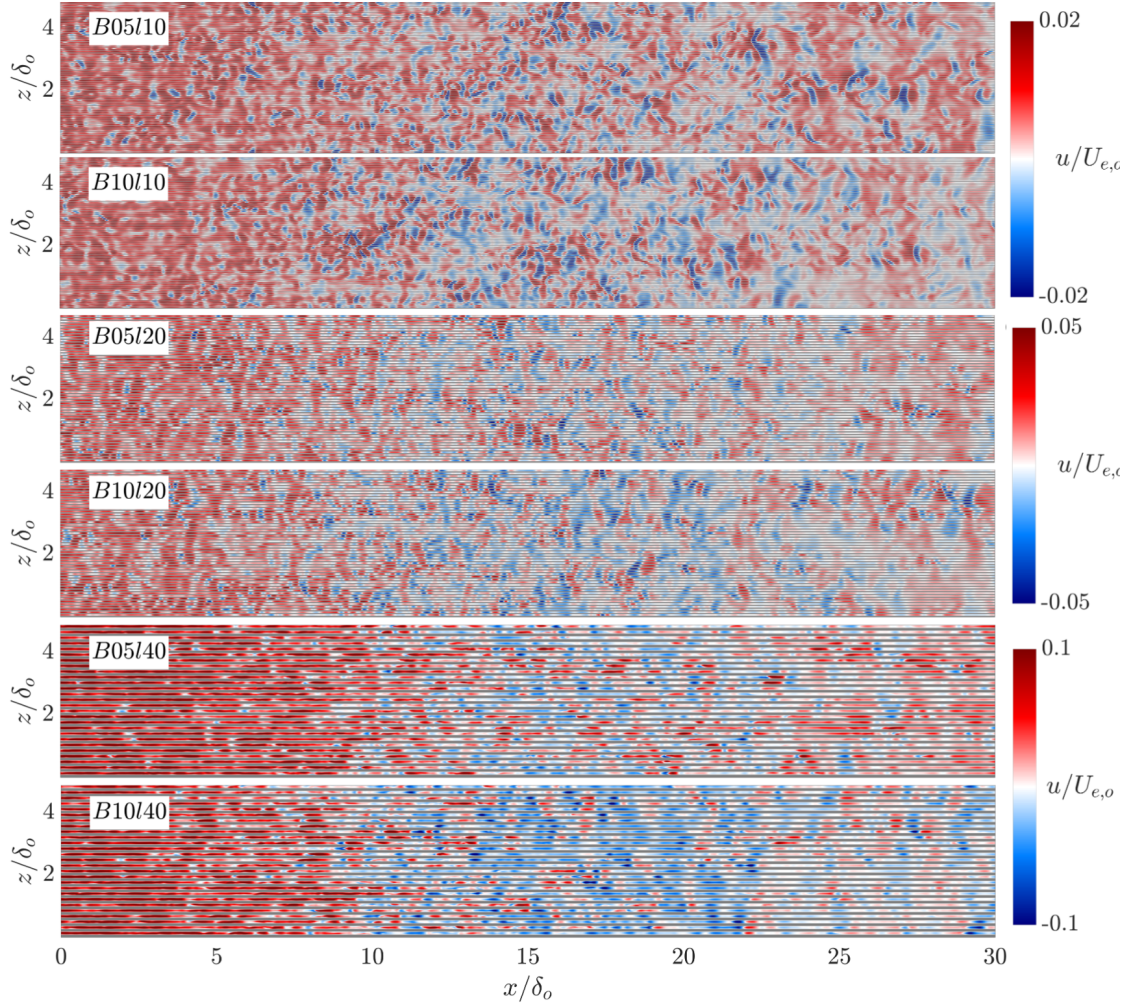

**Fig. 7** Instantaneous streamwise velocity contours in the $x-z$ plane at the half-riblet height. The top two figures are the $l$10 **cases, the middle two are the** $l$20 **cases, and the bottom two are the** $l$40 **cases. Note the different color scaling for the different riblet sizes. Regions occupied by the riblets are shaded in gray.**

### C. Discussion on skin friction modulation

It is evident that the mechanisms responsible for the remarkable drag reduction observed here by riblets in strong APGs are not the same as the drag-reducing mechanisms in equilibrium flows. We now assess potential mechanisms for the highly reduced, and even negative, $\tau_w$. Since riblets do not generate form drag, the skin friction induced by reverse flow within the grooves is the only mechanism that can potentially bring the total drag to zero. Positive drag still arises from the mean shear, $\partial U_c/\partial y$, near the riblet crest. For the net drag to vanish, the reverse flow within the grooves must be strong enough to counteract this (weakened) mean shear. This compensation can occur in two ways: by increasing the magnitude of the reverse flow, and/or by expanding the portion of the groove exposed to reverse rather than forward flow. Our data shows both phenomena play a role. The mean streamwise velocity profiles are compared at various streamwise locations in figure 6. For the riblet cases, the wall-normal coordinate is shifted by the riblet height such that the origin ($y=0$) corresponds to the riblet crest. For all riblet cases, reverse flow beneath the riblet crest is observed at $x/\delta_o=10$ and 20. Larger riblets and stronger APG not only induce a stronger reverse flow within the grooves, but also drive it deeper, increasing the wetted area exposed to the reverse flow. Further discussion will be made in the full paper.

### D. Characterization of KH rollers vortices

To the best of our knowledge, this is the first observation of a *mean* reverse flow beneath riblet grooves that significantly reduces wall shear stress. In APG flows, local and instantaneous regions of reverse flow—resulting in intermittent yet coherent negative wall shear stress—have been reported under certain conditions [33, 48–50]. These events are more prominent than the rare reverse flows caused by extreme ejections in canonical TBLs. Owing to their spanwise coherence, they have been linked to the formation of Kelvin–Helmholtz (KH) roller vortices induced by riblets of specific shapes and sizes. However, riblets which induce a KH instability strong enough to form KH rollers have been shown to increase the mean $\tau_w$ relative to the smooth wall rather than decreasing it, despite local instantaneous regions of $\tau_w < 0$ [13, 14, 33, 48–50]. A reduction in mean $\tau_w$ by KH rollers, as observed in the present study, has not been previously reported. Our results demonstrate that the reverse flow within the riblet grooves is generated by a train of downstream-traveling KH roller vortices. The temporal averaging of these passing vortices yields a mean reverse groove flow. Figures 7 and 8 show the footprints and streamwise pattern of the rollers in the present cases.

Note that empirical quantitative thresholds for the occurrence of KH rollers over riblets have been developed for ZPG flows, along with explanations based on the behavior of transverse flows within the riblet grooves. In general, KH rollers are not typically observed when riblet geometries are within the drag-reducing regime. According to Ref. [48], KH rollers can form over riblets for (approximately) $l^+ \in [15, 40]$, given that the cross-sectional shape is proper to generate strong shear near the riblet crest (see figure 15(c) of the cited study). As mentioned in Section II.B, the riblets are designed in the current study such that the $l10$ cases live below this regime, and the decrease of $u_\tau$ and $l^+$ by the APG should further hinder the formation of KH rollers. The $l20$ cases are designed to be near the low end of the necessary size, and thus $l^+$ should decrease below the KH roller-sustaining regime during the APG. Finally, the $l40$ cases are designed so that $l^+$ transits into the KH roller-sustaining regime during the APG.

Figure 7 exhibits that there are clear spanwise-coherent regions in all cases consisting of alternating forward and reverse flow. Such contours are reminiscent of the regions of instantaneous reverse shear stress in Refs. [48–50], suggesting the presence of KH rollers in the present study. It is further confirmed by instantaneous streamwise velocity in the $x - y$ plane taken in the riblet groove, as provided in figure 8. The KH rollers are readily identified by quasi-circular regions of reverse flow below the riblet crest height. It is evident that for increasing riblet size at a fixed APG, the strength of the KH rollers, and their size increase. The same is true for increasing APG strength at a fixed riblet size.

Recall that the riblet sizes were chosen specifically to examine the occurrence and evolution of the KH rollers. For the small riblets ($l10$, top two rows of figure 7), weak spanwise coherence is observed near the inlet. This agrees with previous findings, as such small riblets are known to support only weak and relatively incoherent KH rollers [14, 48]. However, further downstream, as the APG strengthens and $l^+$ decreases — conditions under which KH rollers are less likely to form according to ZPG-based thresholds and explanations — the rollers not only persist but also grow in coherence and intensity. Our cases show alternating regions of forward and reverse flow with notable spanwise coherence, suggesting that the KH rollers intensify even as $l^+$ decreases under APG. For the $l20$ cases (middle two rows of figure 7), spanwise-coherent structures are evident throughout the entire domain. While the formation of rollers at the ZPG reference location was anticipated, they were expected to weaken downstream as the APG reduces $l^+$, based on findings from ZPG flows. However, similar to the $l10$ cases, both the coherence and the strength of the reverse flow increase in the downstream direction as the APG develops. Compared to the $l10$ cases, the $l20$ cases exhibit stronger reverse flow, greater spanwise coherence, and longer streamwise wavelengths.

Finally, in the $l40$ cases (bottom two rows of figure 7), the signature of KH rollers is not evident at the reference plane as expected. Then, spanwise-coherent regions of intense reverse and forward flow form as the APG grows. While this aligns with expectations based on findings from ZPG flows, the failure of such predictions for the $l10$ and $l20$ cases raises questions about the validity of the explanation in previous studies on the occurrence of the KH rollers. The spanwise coherence is arguably less than the $l20$ riblets. This is likely due to the oversized riblets permitting turbulence to penetrate into the grooves and disrupt the KH rollers, whereas smaller riblets lift near-wall streaks and flow in the grooves is viscous-dominated. However, the intensity of the reverse flow, and the streamwise wavelength of the presumed KH rollers are larger in the $l40$ cases than in the $l10$ and $l20$ cases, thereby explaining the monotonic growth of mean reverse flow with increasing riblet size (refer to figure 6).

### E. Discussion about KH rollers vortices

Our results indicate that the mean reverse groove flow arises from the passage of KH roller vortices, whose lower halves induce a local flow reversal through their negative velocity. Since the bulk flow carrying these vortices continues downstream, the resulting mean reverse flow is, in a sense, a 'pseudo' reversal rather than a true upstream motion. More importantly, the previously established thresholds and explanations for the occurrence of KH rollers may not hold in non-equilibrium TBLs. Moreover, rather than causing drag augmentation, KH roller vortices can actually reduce drag if they are sufficiently strong and prevalent enough to generate a sustained mean reverse flow within the riblet grooves.

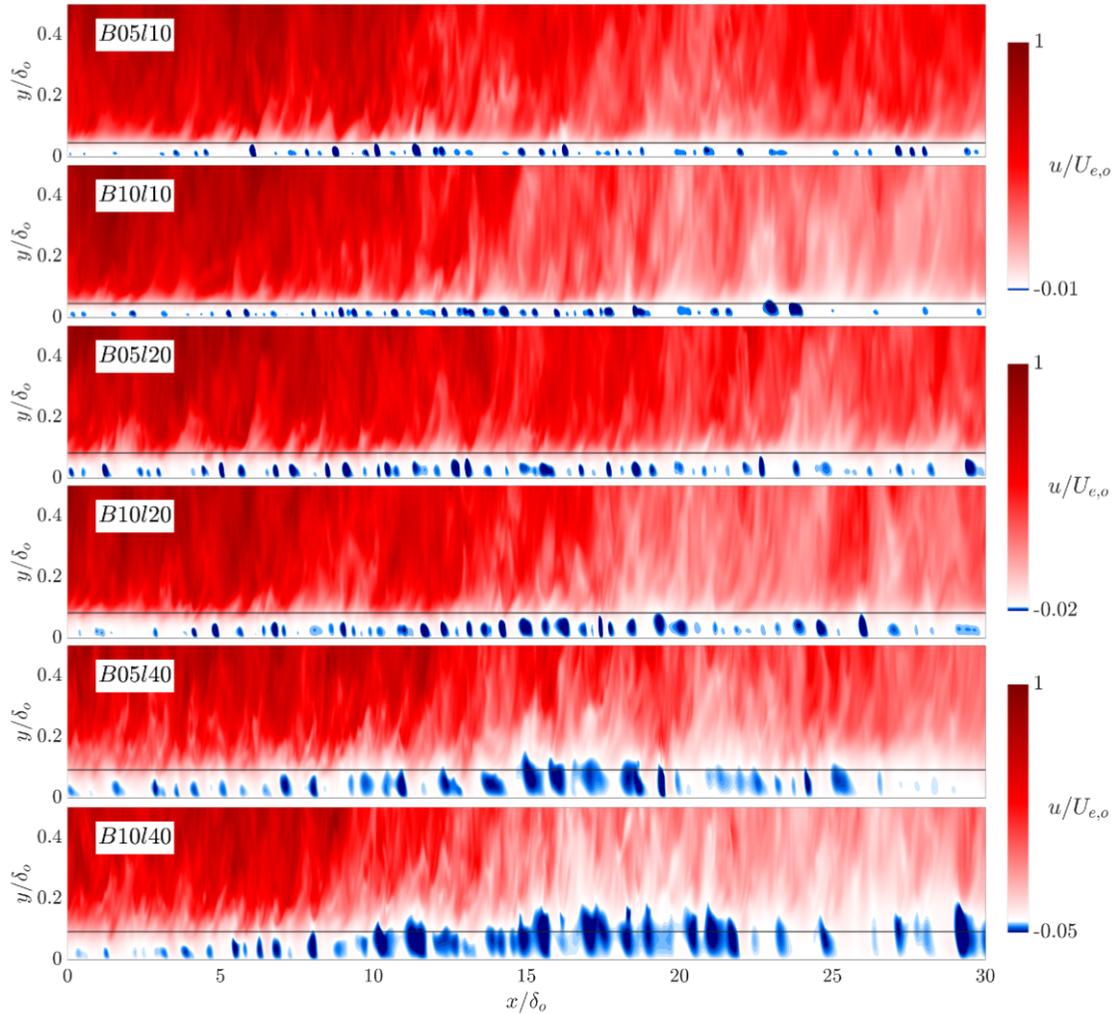

**Fig. 8** Instantaneous streamwise velocity contours in the $x - y$ plane taken at the spanwise location of a riblet valley. The top two figures are the $l10$ cases, the middle two are the $l20$ cases, and the bottom two are the $l40$ cases. The solid horizontal line denotes the height of the riblet crest. Note the varying color scales used for different riblet sizes, as well as the vertical axis stretching applied to enhance the clarity of the near-wall region.

Comparing the cases, it is found that the magnitude of the spanwise-coherent reverse flow increases with APG strength as well as the riblet size, while the proportion of forward flow within the riblet groove decreases (this is most noticeable comparing case $B05l40$ with $B10l40$ in figure 7). It is also interesting to observe the spatial evolution of KH rollers. That is, the streamwise wavelength of the rollers appears to grow with downstream distance. Most notably, in the region of strong APG ($10 \lesssim x/\delta_o \lesssim 25$), the streamwise width of the reverse flow regions tends to grow relative to the upstream quasi-ZPG region. Such spatial development of KH rollers has not been observed in ZPG riblet flows. These observations may indicate a more frequent generation of the KH rollers, stronger KH rollers, or slower convective speed that causes the cluster of consecutive vortices. The full paper will investigate this in more detail.

Previous studies of KH rollers over riblets in ZPG flows attribute their formation to a combination of wall-normal transpiration near the riblet crest (which occurs when the riblets are wider than those typically used for drag reduction) as well as high shear generated at the crest to form an inflection point in the mean velocity profile. This depends on a delicate balance of: 1) sufficiently large riblets to permit transpiration at the crest, and 2) sufficiently small (and sufficiently shaped) riblets such that turbulence does not penetrate into the grooves and high shear is maintained near the crest. In such equilibrium flows, the riblet size in wall units plays a key role, as it governs the balance between the formation of a local inflection point near the crest and the turbulence generated by hairpin vortices. However, this balance does not hold in non-equilibrium APG flows. Our results suggest that rather than being governed by a viscous-scale balance, the APG sustains an internal shear layer of sufficient strength to enable the roll-up of KH vortices, even when the viscous-scaled riblet size would suggest that such structures should not form.

## IV. Conclusion

In this work, direct numerical simulations of adverse pressure gradient (APG) turbulent boundary layers over riblets are performed. The pressure gradient is chosen to lead to significant deceleration but not flow detachment. The objective is to investigate whether riblets maintain their drag-reducing benefit and mechanisms in strongly decelerating TBLs that are representative of flows in engineering settings. Three riblet sizes are examined, allowing them to exist in the drag-reducing, drag-increasing, or KH-roller-inducing regime at the zero-pressure-gradient (ZPG) reference location. Two APG strengths, reaching a maximum $\beta = 5$ and $\beta = 10$ over the smooth wall, are tested. All cases exhibit drag reduction during the APG, and drag reduction grows with both riblet size and APG strength. The drag reduction during the APG is considerably greater than previously reported in ZPG and mild APG flows. Accordingly, the drag curve deviates considerably from the well-established ZPG drag curve, indicating corrections to the ZPG metrics are needed to predict drag modulation when the pressure gradient is non-negligible.

A newly observed feature of the drag curve is a regime of 'forward-force' or thrust production for select combinations of riblet size and APG strength. This is indicative of drag reduction by more than 100%. This forward-force-producing regime, as well as the significant drag reduction in the remaining cases, is a product of mean reverse flow below the riblet crest, despite that massive separation does not occur. The mean reverse flow is shown to be a product of KH roller vortices, which increase in strength, spatial frequency, and size during the strong APG. Ongoing work about the mechanisms and detailed statistics will be included in the full AIAA conference paper.

## Acknowledgments

WW acknowledges the support from AFOSR Grant No. FA9550-25-1-0033, monitored by Dr. Gregg Abate. BSS appreciates the support of NSF GRFP Award No. 2235036. AR acknowledges the support from AFOSR Grant No. FA8655-24-1-7008, monitored by Dr. Douglas Smith. The authors acknowledge the support of the Department of Defense Carpenter Supercomputing Center and San Diego Supercomputing Center for providing computational resources.